\documentclass[
reprint,
superscriptaddress,
nofootinbib,
amsmath,amssymb,
aps,
prb,
twocolumn,
longbibliography,
]{revtex4-2}
\usepackage[utf8]{inputenc}
\usepackage{amsfonts}
\usepackage{amssymb}
\usepackage{amsmath}
\usepackage{xcolor}
\usepackage{bbold}
\usepackage{subcaption}
\usepackage[margin=0.6in]{geometry}

\usepackage{graphicx}
\date{\today}

\makeatletter
\let\Hy@backout\@gobble
\makeatother

\begin{document}
\title{Spin-orbit-assisted electron pairing 
in 1D waveguides}
\author{Fran\c{c}ois Damanet}
\author{Elliott Mansfield}\affiliation{Department of Physics and SUPA, University of Strathclyde, Glasgow  G4 0NG, Scotland, United Kingdom}
\affiliation{Pittsburgh Quantum Institute, Pittsburgh, PA 15260, USA}
\author{Megan Briggeman}
\author{Patrick Irvin}
\author{Jeremy Levy}
\affiliation{Department of Physics and Astronomy, University of Pittsburgh, Pittsburgh, PA 15260, USA}
\affiliation{Pittsburgh Quantum Institute, Pittsburgh, PA 15260, USA}
\author{Andrew J. Daley}
\affiliation{Department of Physics and SUPA, University of Strathclyde, Glasgow  G4 0NG, Scotland, United Kingdom}
\affiliation{Pittsburgh Quantum Institute, Pittsburgh, PA 15260, USA}

\begin{abstract}

Understanding and controlling the transport properties of interacting fermions is a key forefront in quantum physics across a variety of experimental platforms. Motivated by recent experiments in 1D electron channels written on the $\mathrm{LaAlO_3}$/$\mathrm{SrTiO_3}$ interface, we analyse how the presence of different forms of spin-orbit coupling (SOC) can enhance electron pairing in 1D waveguides. We first show how the intrinsic Rashba SOC felt by electrons at interfaces such as $\mathrm{LaAlO_3}$/$\mathrm{SrTiO_3}$ can be reduced when they are confined in 1D. Then, we discuss how SOC can be engineered, and show using a mean-field Hartree-Fock-Bogoliubov model that SOC can generate and enhance spin-singlet and triplet electron pairing. Our results are consistent with two recent sets of experiments~[Briggeman \textit{et al.}, arXiv:1912.07164; Sci.\ Adv.\ \textbf{6}, eaba6337 (2020)] that are believed to engineer the forms of SOC investigated in this work, which suggests that metal-oxide heterostructures constitute attractive platforms to control the collective spin of electron bound states. However, our findings could also be applied to other experimental platforms involving spinful fermions with attractive interactions, such as cold atoms.

\end{abstract}
\maketitle

\section{Introduction}

There is a fundamental interest in investigating transport in reduced-dimensionality systems, where interactions play a prominent role and can lead to exotic phases of matter~\cite{Giamarchi2003book}. Understanding transport dynamics in these systems and how they can be controlled is also crucial to shed light on the properties and the further development of useful materials. 

In the solid-state, different platforms have been developed to offer playground for the investigation of strongly-correlated systems. Of particular interest for this work are the metal-oxide heterostructures, such as the $\mathrm{LaAlO_3}$ and $\mathrm{SrTiO_3}$ interface (LAO/STO)~\cite{ohtomo_high-mobility_2004, Pai_2018}. By depositing just a few layers of LAO onto the bulk STO substrate, a 2D electron gas (2DEG) can be formed at the interface. This 2DEG is highly controllable, and can be engineered to display many interesting phenomena, such as superconductivity~\cite{Reyren1196}, tuneable transport~\cite{Thiel2006}, Rashba spin-orbit coupling~\cite{Caviglia2010, Shalom2010}, or ferromagnetic phases~\cite{Brinkman2007}. Notably, using conductive atomic-force microscope (c-AFM) lithography, it has been shown that nanostructures at the LAO/STO interface can be created, allowing in particular for the exploration of quantum transport of electrons with tunable attractive interactions~\cite{Cheng2016} in engineered quasi-1D nanowires~\cite{cen_nanoscale_2008}. This method has led to the observation of quantized ballistic transport of single and paired electrons~\cite{Annadi2018}, as well as more exotic bound states of three or more electrons~\cite{Briggeman2020Science}. The AFM tip used to create the nanostructures acts as a nanoscale pencil that can reversibly tune the transport properties of the interface from insulating to conducting via (de)protonation~\cite{Bi2016,Brown2016}, and offers great potential to study and engineer new transport phenomena and new phases of quantum matter. The level of control being developed here evokes comparison with the high control of quantum simulators in other platforms~\cite{cirac_goals_2012,Lanyon2010Towards}, such as cold atoms~\cite{bloch_quantum_2012}, coupled light resonator arrays~\cite{grujic_non-equilibrium_2012}, trapped ions~\cite{blatt_quantum_2012}, superconducting circuits~\cite{houck_-chip_2012}, or Rydberg atoms~\cite{Browaeys2020Manybody}. 

Driven by the long-term prospect of developing analog quantum simulators in the solid-state, more recent experiments in LAO/STO interfaces have shown that by spatially modulating the AFM tip during the writing process, it is possible to create laterally undulating wires~\cite{briggeman2019lateral} and 1D Kronig-Penney-like superlattice structures~\cite{briggeman2019vertical}. Compared to straight waveguides~\cite{Annadi2018}, these devices exhibit stable fractional conductance plateaux and enhanced pairing of electrons.
A possible explanation for some of the observed phenomena is that the modulation engineers a spin-orbit coupling (SOC) in the waveguide which modifies its transport properties. SOC constitutes a useful resource to manipulate spins in a wide range of applications and has led to the discovery of new topological classes of materials~\cite{Manchon2015New}. SOC appears naturally in crystals that lack an inversion symmetry, such as the LAO/STO interface, but could also be engineered artificially by creating effective broken inversion symmetries via applied electric fields.

The goal of this theoretical work is to analyse how the presence of spin-orbit interactions can enhance the pairing of electrons in 1D waveguides such as the ones realized on the LAO/STO interface, elaborating on the theory discussed in~\cite{briggeman2019vertical}. Our work suggests that heterostructures constitute an attractive platform for controlling collective spin states of electron pairs. However, the theory presented here is quite general and could be used to describe other experimental platforms involving 1D systems of spinful fermionic particles with attractive interactions and SOC, such as cold atoms~\cite{Lebrat2018PRX, Krinner2017}. SOC can indeed be induced in these systems via the use of artificial gauge fields~\cite{Dalibard2015Introduction,Goldman2014Light,Farias2014Degenerate,Wang2018Dirac}, and the interplay between SOC and interactions is starting to be explored with cold atoms~\cite{Cheuk2012,Liu2009,Wang2012,DellAnna2011,DellAnna2012}, showing notably the possibility to enhance pairing via SOC. Our findings are consistent with these results but in different forms and parameter regimes.

This paper is structured as follows. In Sec.~II, we review the band model typically used to describe the origin of the intrinsic Rashba spin-orbit coupling that can exist in crystals that lack of inversion symmetry, such as the LAO/STO interface, and provide some arguments on how it can be reduced when the electrons are confined in a 1D channel. We then elaborate on ways to engineer different forms of SOC in the waveguide. In Sec.~III, we derive a single-particle model for these electrons that includes the SOC and the electron-electron interactions at the mean-field level. In Sec.~IV, we solve our model to study the interplay between interactions and SOC, showing that the SOC can enhance spin-singlet and spin-triplet pairing. In Sec.~V, we finally conclude and present some perspectives of our work.

\section{Electrons at interfaces with SOC}

In this section, we first present a review of a simple band model for electrons at heterostructure interfaces that explains the origin of the Rashba spin-orbit coupling, following~\cite{ZhongPRB13, Kim13}. Then, elaborating on the model presented in~\cite{Kim13}, we show how it can be reduced when the electrons are confined in a 1D channel. Finally, we discuss ways to engineer different forms of SOCs in such systems.

For concreteness and clarity, the theory presented below is based on the electronic structure of the LAO/STO interface. However, the theory is quite general, and could be applied to other interfaces by repeating the procedure below with modified parameters and initial electron orbitals, depending on where the Fermi energy lies in the relevant material.

\subsection{Band structure of electrons at an interface with SOC: a short review for LAO/STO}

\begin{figure*}
    \centering
    \includegraphics[width=\textwidth]{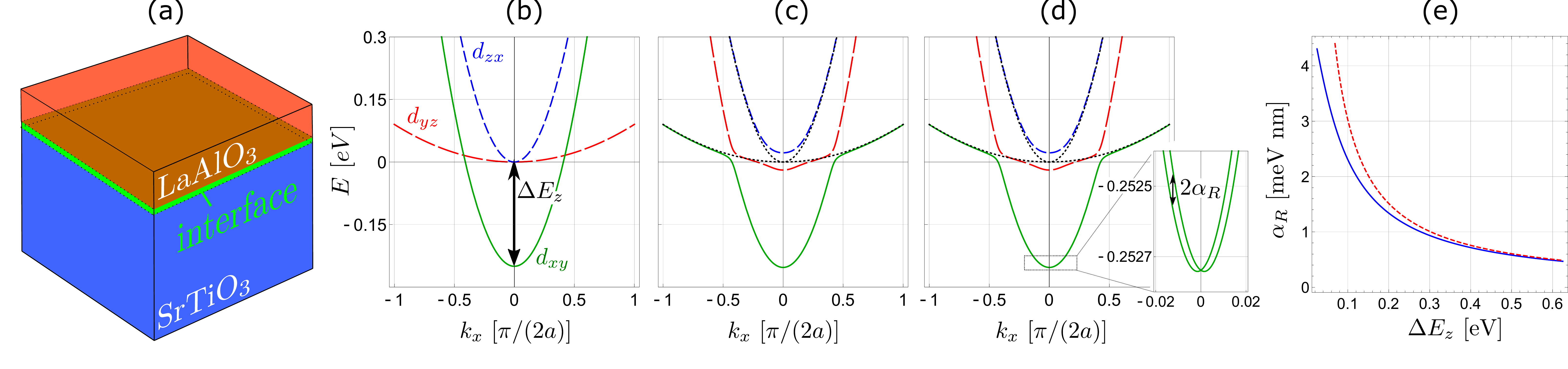}
    \caption{\small{\textbf{Electrons at the 2D LAO/STO interface.} (a) Diagram of a typical LAO/STO interface (green) between a few layers of $\mathrm{LaAlO_3}$ (red) on top of bulk $\mathrm{SrTiO_3}$ (blue). (b) Energies of $H_0$ [Eq.~(\ref{H0})] as a function of $k_x \in [-\pi/(2a),\pi/(2a)]$ where $a = 0.392$ nm is the lattice spacing. The parameters are $m_h = 6.8 m_e$, $m_l = 0.41 m_e$, $\Delta E_z = 0.25$ eV, where $m_e$ is the electron mass. The spin-degenerate orbital $d_{xy}$ (green solid line) is the lowest energy state due to the confinement energy $\Delta E_z$ along $z$, compared to $d_{yz}$ (red long-dashed line) and $d_{zx}$ (blue short-dashed line). (c) Energies of $H_0 + H_\mathrm{aso}$ as a function of $k_x$, for $\Delta_{aso} = 19.3$ meV and other parameters as above. The atomic spin-orbit coupling Hamiltonian $H_\mathrm{aso}$ [Eq.~(\ref{Haso})] mixes the orbitals at the crossing points, so does not affect much the low-energy electrons in the bottom of the lowest orbital $d_{xy}$ (near $k_x \approx 0$). The dotted black lines correspond to the energies of $H_0$ as in Fig.~1b for comparison. (d) Energies of the full Hamiltonian $H_\mathrm{tot} = H_0 + H_\mathrm{aso} + H_a$ [Eq.~(\ref{Htot})] as a function of $k_x$, for $\Delta_z = 20$ meV and other parameters as above. The Hamiltonian $H_a$ lifts the spin degeneracy and generates a linear Rashba SOC of strength $\alpha_R$ for the electrons in the lowest band, as highlighted in the inset. (e) Rashba SOC strength as a function of $\Delta E_z$ obtained from the numerical diagonalization of $H$ [Eq.~(\ref{Htot})] (blue solid line) and from the perturbative theory [Eq.~(\ref{alpharP})] (dashed red line)], for other parameters as above~\cite{Comment1}. The Rashba SOC decreases for increasing confinement along $z$.
}}
    \label{fig1}
\end{figure*}

Figure~\ref{fig1}(a) presents a sketch of a typical LAO/STO structure, where a few layers of $\mathrm{LaAlO_3}$ are placed on the top of bulk $\mathrm{SrTiO_3}$. In STO-based interfaces, the Fermi energy lies in the 3$d$ $t_{2g}$ orbitals $d_{yz}$, $d_{xz}$, and $d_{xy}$ of the Ti ions near the interface. The electrons are naturally confined in the direction normal to it (labelled as the $z$-direction) to form a two-dimensional electron gas [green layer in Fig.~\ref{fig1}(a)]. The backbone of the electron dynamics can be accounted for by a simple six-dimensional Hamiltonian $H_0$, which, in the orbital and spin basis $(d_{yz},d_{xz}, d_{xy}) \otimes (\uparrow, \downarrow)$, takes the form~\cite{Kim13, Bistritzer2011, Joshua2012}
\begin{multline}\label{H0}
H_0 = \\ \small{\begin{pmatrix}
\frac{\hbar^2 k_x^2}{2 m_h} + \frac{\hbar^2 k_y^2}{2 m_l} & 0 & 0 \\
0 & \frac{\hbar^2 k_x^2}{2 m_l} + \frac{\hbar^2 k_y^2}{2 m_h}  & 0 \\
0 & 0 & \frac{\hbar^2 k_x^2}{2 m_l} + \frac{\hbar^2 k_y^2}{2 m_l}  - \Delta E_z
\end{pmatrix}} \otimes \mathbb{1}_{2},
\end{multline}
where $m_h$ and $m_l$ are effective heavy and light masses, $\Delta E_z$ is the energy splitting due to the natural confinement of the bands along $z$ -- making $d_{xy}$ the lowest band at small $k$ values -- and $\mathbb{1}_{2}$ is the identity operator acting on the two-dimensional spin Hilbert space. The Hamiltonian $H_0$ is naturally diagonal in the orbital basis and spin degenerate. Figure~\ref{fig1}(b) shows its energies as a function of $k_x$ for reasonable parameter values.

The Hamiltonian $H_0$ is too simple to capture all the interesting features of the electron gas. In order to account for the effect of the atomic spin-orbit coupling, the Hamiltonian
\begin{equation}\label{Haso}
H_{\mathrm{aso}} \propto \mathbf{L} \boldsymbol{\cdot} \boldsymbol{\sigma} = i \Delta_\mathrm{ASO} \begin{pmatrix}
0 &  \sigma_z & -  \sigma_y \\
-  \sigma_z & 0 &  \sigma_x \\
 \sigma_y & - \sigma_x & 0 \\
\end{pmatrix},
\end{equation}
is added to $H_0$, where $\mathbf{L} =  \mathbf{r} \times \mathbf{p}$ is the orbital momentum operator, $\boldsymbol{\sigma} = ( \sigma_x, \sigma_y, \sigma_z)$ is the vector of Pauli operators, and $\Delta_\mathrm{aso}$ is the atomic spin-orbit coupling strength. Figure~\ref{fig1}(c) shows the energies of $H_0 + H_\mathrm{aso}$ as a function of $k_x$, where $H_\mathrm{aso}$ has the effect to mix the eigenstates of $H_0$. We chose $\Delta_\mathrm{aso} = 19.3$ meV as in~\cite{ZhongPRB13, Bistritzer2011}, motivated by the fact that it leads to the same modifications of the orbitals as the ones obtained via a more refined calculation of the band structure via density functional theory (DFT)~\cite{Bistritzer2011}.

Finally, due to the broken inversion symmetry at the interface along $z$, an additional coupling of the orbital $d_{xy}$ to $d_{yz}$ and $d_{xz}$ appears, at the origin of a Rashba spin-orbit coupling, as discussed below. This effect can be accounted for via a third Hamiltonian of the form~\cite{ZhongPRB13,  Kim13,LaShell1996,Petersen2000}
\begin{equation}\label{Ha}
H_a = i \Delta_z a \begin{pmatrix}
0 & 0 &  k_x \\
0 & 0 &  k_y \\
-  k_x & - k_y & 0 \\
\end{pmatrix} \otimes \mathbb{1}_{2},
\end{equation}
where $a = 0.392$ nm is the lattice spacing and $\Delta_z$ is the overall energy scale. Figure~\ref{fig1}d shows the eigenvalues of the full Hamiltonian 
\begin{equation}\label{Htot}
H_\mathrm{tot} = H_0 + H_\mathrm{aso} + H_a,
\end{equation} 
using $\Delta_z = 20$ meV, as can be again extracted from DFT~\cite{ZhongPRB13}. As can be seen in the inset of the figure, $H_a$ causes a lifting of the spin degeneracy of the two lowest energies. For small $k$, this energy splitting has the form of a linear Rashba term
\begin{equation}
\Delta E_R = \alpha_R (k_x \pm i k_y),
\end{equation}
with $\alpha_R$ being the spin-orbit coupling strength (with dimensions of energy $\times$ length). The value of $\alpha_R$ can be obtained from the model by simply fitting the difference between the two lowest energies of $H_\mathrm{tot} = H_0 + H_\mathrm{aso} + H_a$ at small momentum $\mathbf{k}$ by a linear function of $k_x$ (or $k_y$); the slope of it providing $2 \alpha_R$. Alternatively, one can write $H_a$ in the basis of the eigenstates of $H_0(\mathbf{k} = 0) + H_ \mathrm{aso}$ and extract $\alpha_R$ from its matrix form, after identifying the off-diagonal elements corresponding to the coupling between the two lowest bands. Finally, since typically we have $\Delta E_z \gg \Delta_z, \Delta_\mathrm{aso}$, one can use a second order perturbation theory (first order in $H_\mathrm{aso}$ and $H_a$) to obtain the following analytical expression~\cite{Kim13, ZhongPRB13}
\begin{equation}\label{alpharP}
\alpha_R = 2 a \frac{\Delta_z \Delta_\mathrm{aso}}{\Delta E_z},
\end{equation}
valid for $\Delta E_z \gg \Delta_z, \Delta_\mathrm{aso}$. Experimentally, the value $\alpha_R$ is usually found to be around $1$-$5$ meV$\,$nm~\cite{Caviglia2010,Shalom2010}, and the model presented above predicts compatible values of $\alpha_R$. Figure~\ref{fig1}(e) shows $\alpha_R$ obtained from both the numerical diagonalization of $H_\mathrm{tot}$ and the perturbative theory~(\ref{alpharP}) as a function of the confinement energy $\Delta E_z$ and for other parameters taken from~\cite{ZhongPRB13}. As can be seen in Fig.~\ref{fig1}(e), increasing the confinement along $z$ reduces the linear Rashba spin-orbit coupling felt by the low-energy electrons located at the bottom of the two lowest bands.

\subsection{Effects of confinement in 1D}

\begin{figure}
    \centering
    \includegraphics[width=0.475\textwidth]{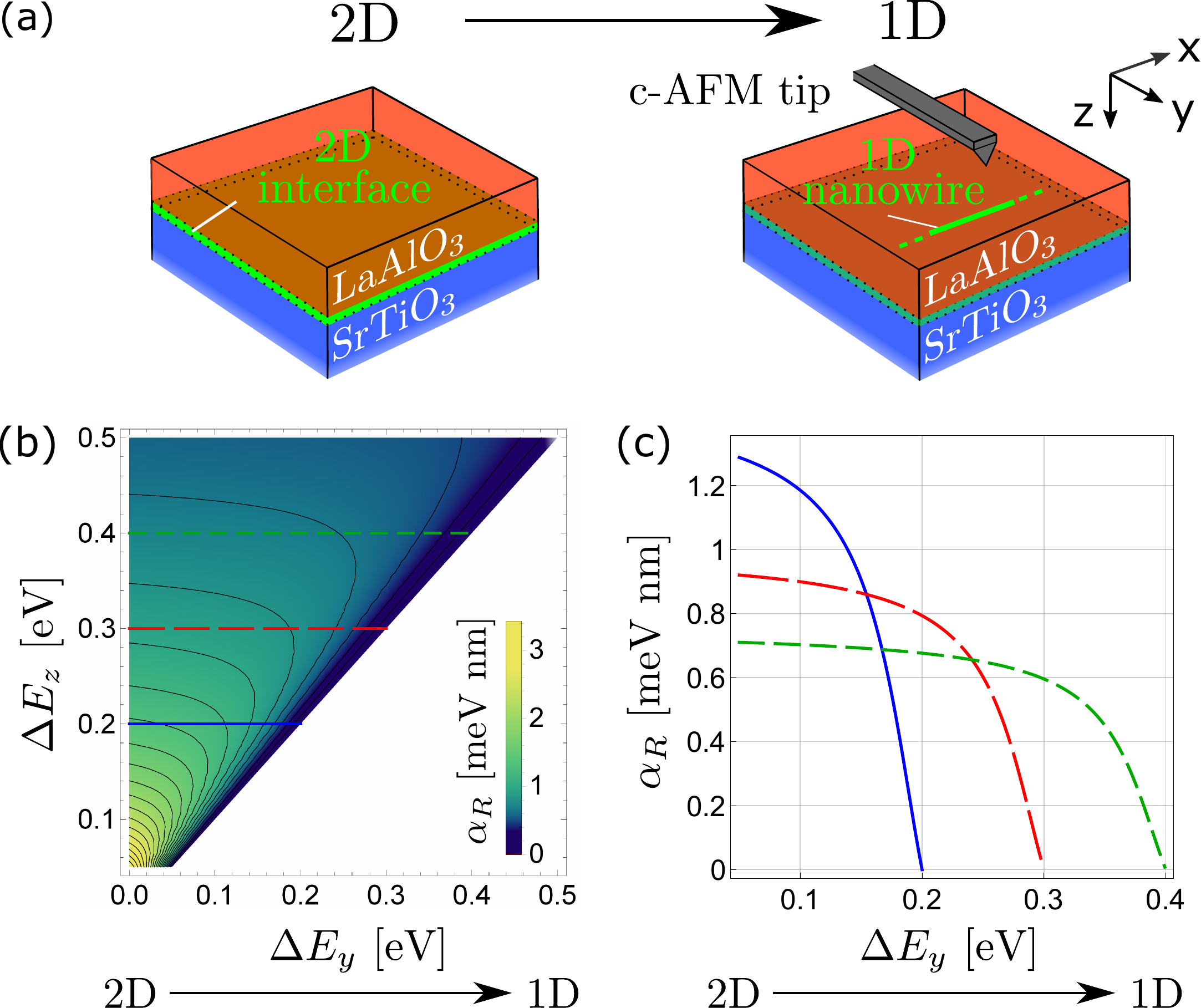}
    \caption{\small{\textbf{Rashba SOC from 2D to 1D.} (a) Sketch of the transition from a 2D conducting interface to 1D nanowire as realized via c-AFM lithography. (b) Rashba SOC strength $\alpha_R$ as a function of the lateral and vertical confinements $\Delta E_y$ and $\Delta E_z$. The values of $\alpha_R$ along the horizontal lines (corresponding to three specific values of $\Delta E_Z$) are shown more clearly on panel (c). (c) Rashba SOC strength $\alpha_R$ as a function of the confinement $\Delta E_y$ for $\Delta E_z = 0.2$ eV (solid blue line), $0.3$ eV (red long-dashed line) and $0.4$ eV (green short-dashed line). Increasing the confinement along $\Delta E_y$ reduces the dimensionality of the system and the Rashba SOC.}}
    \label{fig2}
\end{figure}

We saw in the previous section that the vertical confinement of the electrons can be accounted for by a phenomenological parameter $\Delta E_z$ in $H_0$ [Eq.~(\ref{H0})]. As suggested in~\cite{Kim13}, it is reasonable to think that the lateral confinement felt by the electrons in a 1D waveguide [as can be realized via c-AFM lithography as sketched in Fig~\ref{fig2}(a)] could be modeled in a similar way. Labelling $x$ as the direction of the waveguide, the modified $H_0$ reads~\cite{Kim13}
\begin{multline}\label{H02}
H_0' = \\ \small{\begin{pmatrix}
\frac{\hbar^2 k_x^2}{2 m_h} + \frac{\hbar^2 k_y^2}{2 m_l} & 0 & 0 \\
0 & \frac{\hbar^2 k_x^2}{2 m_l} + \frac{\hbar^2 k_y^2}{2 m_h} - \Delta E_y & 0 \\
0 & 0 & \frac{\hbar^2 k_x^2}{2 m_l} + \frac{\hbar^2 k_y^2}{2 m_l}  - \Delta E_z
\end{pmatrix}} \otimes \mathbb{1}_{2},
\end{multline}
where $\Delta E_y$ is the engineered confinement along $y$. The presence of $\Delta E_y$ lifts the degeneracy between the $d_{xz}$ and the $d_{yz}$ bands.

By diagonalizing 
\begin{equation}\label{Htotprime}
H_\mathrm{tot}' = H_0' + H_\mathrm{aso} + H_a
\end{equation} 
and then extracting numerically the value of the Rashba spin-orbit coupling strength $\alpha_R$ along the direction of the nanowire $k_x$ as explained before, we can show that $\alpha_R$ decreases monotically as a function of $\Delta E_y$ and goes to zero for $\Delta E_y  = \Delta E_z$, when the bands $d_{xz}$ and $d_{xy}$ become degenerate. This can be seen in Fig.~\ref{fig2}(b), showing $\alpha_R$ as a function of both $\Delta E_z$ and $\Delta E_y$ for a wide range of reasonable values with $\Delta E_y \leqslant \Delta E_z$ so that the orbital $d_{xy}$ remains the ground state, degenerate with $d_{zx}$ only when $\Delta E_y = \Delta E_z$. Figure~\ref{fig2}(c) shows more clearly three line cuts of Fig.~\ref{fig2}(b), i.e.,  $\alpha_R$ as a function of $\Delta E_y$ for $\Delta E_z = 0.2$, $0.3$ and $0.4$ eV. While precise values of $\Delta E_y$ and $\Delta E_z$ are difficult to estimate, this result suggests that in quasi-1D nanowires, the intrinsic Rashba spin-orbit coupling due to the broken vertical inversion symmetry could be significantly smaller than in the 2D case (i.e., when $\Delta E_y = 0$). Let us emphasize again that the procedure described here to study the impact of the confinement and obtain these findings is quite general and can easily be adapted to other materials.

\subsection{Engineering spin-orbit couplings}

The model presented in the previous section suggests that the usual Rashba SOC at interfaces with broken inversion symmetry could be smaller when the electrons are confined in quasi-1D nanostructures. We discuss here ways to recover a significant value of Rashba SOC, as well as other forms of SOC, before studying its effects on the transport properties of the electrons.

As mentioned in the introduction, recent transport experiments in a 1D modulated waveguide have shown signatures of SOC, both directly~\cite{briggeman2019lateral} and potentially through observation of spin-orbit enhanced electron pairing~\cite{briggeman2019vertical}, which we present a description of in Sec.~IV. The qualitative argument is the following: assume the electrons are confined along $x$. When they travel through  the waveguide with a velocity $\mathbf{v} = v_x \mathbf{e}_x$ wher $\mathbf{e}_i$ is a unit vector along $i = x,y,z$, they feel an electric field produced by the modulation. For a vertical modulation as in~\cite{briggeman2019vertical}, it would have the form $\mathbf{E}_\mathrm{eff} \propto E_\mathrm{eff}(x) \mathbf{e}_z$. This would lead to an effective magnetic field $\mathbf{B}_\mathrm{so} \propto \mathbf{v} \times \mathbf{E}_\mathrm{eff}$ along $y$ and then produce an energy shift of the form $\propto \boldsymbol{\sigma}\boldsymbol{\cdot}\mathbf{B}_\mathrm{so}$. By contrast, a lateral modulation as in~\cite{briggeman2019lateral} would be associated with an electric field $\mathbf{E}_\mathrm{eff} \propto E_\mathrm{eff}(x) \mathbf{e}_y$ and thus a magnetic field along $z$. This reasoning is of course valid for applied external fields that are not necessarily position-dependent. This is at the root of the control of the Rashba SOC at the LAO/STO interface~\cite{Caviglia2010,Shalom2010}.

Hence, despite the fact that the confinement along $y$ can drastically reduce the Rashba SOC at interfaces with natural broken inversion symmetry, it is possible to engineer it in different forms depending on the direction of applied electric fields, resulting e.g.~from the design of side-gate voltages in specific configurations. It is thus experimentally relevant to explore the effects of SOC on the transport dynamics of electrons, which is the purpose of the next sections.

\section{Model for electron transport in 1D waveguides with SOC}

Having looked at how to control artificial SOC in electron waveguides, in this section, we now present a model to describe the interplay between interactions and different forms of SOC on the transport properties of the electrons. We present successively the single-particle basis of our model (Sec.~III.~A), the electron-electron interactions we consider (Sec.~III.~B), and finally how to solve the relevant equations of motion (Sec.~III.~C), by means of a self-consistent mean-field model.

\subsection{Single-particle model with SOC}

We assume the electrons are confined in a 1D channel along $x$, as depicted e.g. in Fig.~\ref{fig1}(a), and that an external out-of-plane magnetic field $B$ is applied along $z$, introducing a Landau quantization effect~\cite{BEENAKKER19911}. The waveguide (typically $\sim 50-1000$ nm long in LAO/STO devices) is connected at both ends to unbiased leads that act as reservoirs (not shown). We consider the case where the Fermi energy of the electrons in the waveguide is tunable via a gate voltage but remains close to the bottom of the lowest band of $H_\mathrm{tot}'$ [Eq.~(\ref{Htotprime})], so that the electrons have too-low momenta $k_x$ to populate other bands. For LAO/STO interfaces, this is motivated by the fact that in conductive nanostructures created via c-AFM lithography, the typical carrier density is around  $n \sim 0.5$-$1 \times 10^{13}$ cm$^{-2}$~\cite{Bi2016}. Roughly speaking this would mean an associated momentum $k = \sqrt{2\pi n} \sim 0.7$-$1 \times 10^{-6} \,(\pi/a)$, i.e., far from the crossing point between the $d_{xy}$ and $d_{yz}$ orbitals [around $\sim 0.25 \,(\pi/a)$ in Fig.~\ref{fig1}(b-d)]. Hence, all the transport channels are assumed to originate from a single band. 

We generalize the single-particle model described in detail in~\cite{Annadi2018} to include spin-orbit couplings. The Hamiltonian for spin up and down electrons in Landau gauge where the vector potential reads $\mathbf{A} = (-B y,0 ,0)$ takes the form
\begin{multline}\label{Hwaveguide}
H = \frac{(p_x - e B y)^2}{2 m_x} + \frac{p_y^2}{2 m_y} + \frac{p_z^2}{2 m_z} + V(y) + V(z) - \mu  \\- g \frac{\mu_B}{2} B \sigma_z + \frac{\left( \alpha_v \sigma_y + \alpha_l \sigma_z\right)}{\hbar}\left(p_x - eB y\right).
\end{multline}
The first line describes the (spin-degenerate) kinetic and potential energies of the electrons, where $p_i$ and $m_i$ ($i= x,y,z$) are the electron momentum operator and effective mass components along the different directions, where 
\begin{equation}\label{potential}
\begin{aligned}
V(y) = \frac{m_y \omega_y^2 y^2}{2}, \quad V(z) = \begin{cases} \frac{m_z \omega_z^2 z^2}{2} & \mathrm{for }\; z \geqslant 0 \\
+\infty & \mathrm{for }\;  z < 0, 
\end{cases}
\end{aligned}
\end{equation}
are parabolic and half-parabolic potentials describing the transverse confinement along $y$ and $z$, with $\omega_y$ and $\omega_z$ the trapping frequencies, and where $\mu$ is the chemical potential. Here, we consider indeed that electron cannot penetrate into one of the layer (as it is the case for the LAO layer in LAO/STO interfaces), i.e., the confining potential along $z$ is infinite for $z < 0$. The second line of Eq.~(\ref{Hwaveguide}) corresponds to the Zeeman energy term due to the out-of-plane magnetic field $B$, with $g$ the Landé factor and $\mu_B$ the Bohr magneton, and the last two terms are effective SOC terms of strength $\alpha_v$ and $\alpha_l$. As explained in the previous section, these two terms could be engineered through applications of electric fields vertically ($v$) or laterally ($l$), and thus referred as vertically-induced or laterally-induced SOC. Note that in the case of modulated electric fields, the SOCs should in principle be spatially dependent, which would drastically complicate the model. In this specific scenario, we make the approximation that $\alpha_v$ and $\alpha_l$ describe the root-mean-square values of the varying parameters. 

Since the Hamiltonian~(\ref{Hwaveguide}) is translationally invariant along $x$, we can make the replacement $p_x \to \hbar k_x \equiv \hbar k$. The $x$ and $y$ motional part of Eq.~(\ref{Hwaveguide}) can then be decoupled using
\begin{equation}\label{Hmotional}
\begin{aligned}
&\frac{(\hbar k - e B y)^2}{2 m_x} + V(y) + \frac{\left( \alpha_v \sigma_y + \alpha_l \sigma_z\right)}{\hbar} \left(\hbar k - eB y\right) \\
&= \frac{\hslash^2 k^2}{2m_x}\frac{\omega_y^2}{\Omega^2} +  \frac{m_y \Omega^2}{2}\left(y - y_0(k) \right)^2 - \frac{e^2 B^2}{2 m_y \Omega^2\hbar^2} \left(\alpha_l^2 + \alpha_v^2 \right)\\
&\quad +\frac{\omega_y^2}{\Omega^2} (\alpha_v \sigma_y + \alpha_l\sigma_z) k
\end{aligned}
\end{equation}
where
\begin{align}
\omega_c &=\frac{eB}{\sqrt{m_x m_y}}, \\
\Omega &= \sqrt{\omega_y^2 + \omega_c^2 }, \\ \label{COM}
y_0(k) &= \frac{\hbar \omega_c k}{\sqrt{m_y m_x} \Omega^2} + \frac{e B}{\hbar m_y \Omega^2} \left(\alpha_v \sigma_y + \alpha_l \sigma_z\right), 
\end{align}
are respectively the cyclotron frequency, the effective frequency and the momentum-and-spin-dependent center of the lateral trapping. If the second term of Eq.~(\ref{COM}) was not present, then the eigenstates of the motional part of~(\ref{Hwaveguide}) would correspond to plane waves $|k\rangle$ along $x$ and 2D harmonic oscillator eigenstates $|m,n\rangle$ along $y$ and $z$, where $m, n \in \mathbb{N}$ are the quantum numbers labelling them. In this case, the position $y_0(k)$ of the center-of-mass of the harmonic oscillator eigenstate along $y$ depends on the momentum $k$ of the electron along $x$. However, due to the last term of Eq.~(\ref{COM}), this center-of-mass position is in principle also spin-dependent, which complicates this picture. In Appendix A, we show however that for reasonable values of the parameters, the second term of Eq.~(\ref{COM}) can be neglected. We will thus ignore the spin-dependence of $y_0(k)$ in the following, as well as the term quadratic in $\alpha_v$ and $\alpha_l$ in Eq.~(\ref{Hmotional}), since it has the same order of magnitude. In the basis $\{|m,n,k,\sigma\rangle\}$ ($\sigma = \uparrow, \downarrow$), the full Hamiltonian~(\ref{Hwaveguide}) has thus the matrix elements
\begin{multline}
\langle m,n,k | H | m,n,k\rangle \simeq \\ \begin{pmatrix}
E_{mnk} - \frac{g \mu_B B}{2} + \alpha_l \frac{\omega_y^2}{\Omega^2} k  & - i \alpha_v \frac{\omega_y^2}{\Omega^2} k \\
  i \alpha_v \frac{\omega_y^2}{\Omega^2} k & E_{m,n,k} + \frac{g \mu_B B}{2} -\alpha_l \frac{\omega_y^2}{\Omega^2} k
\end{pmatrix}, 
\end{multline}
where
\begin{equation}\label{Emnk}
E_{mnk} = \frac{\hslash^2 k^2}{2m_x}\frac{\omega_y^2}{\Omega^2} +  \hslash \Omega \left(m + \frac{1}{2}\right) + \hslash \omega_z\left(2n + \frac{3}{2}\right) - \mu 
\end{equation}
are the energies of the states $|m,n,k\rangle$. In the absence of SOC (i.e., for $\alpha_v = \alpha_l = 0$), the states $|m,n,k,\sigma\rangle$ are the eigenstates of $H$ with energies 
\begin{equation}
\xi_{mn\sigma k} =  E_{mnk} - s(\sigma) g \mu_B B,
\end{equation} 
where $s(\downarrow) = -1/2$ and $s(\uparrow) = 1/2$. The presence of an applied lateral or vertical electric field respectively alters the Zeeman splitting energy or mixes the different spin species within a given transverse mode $|m,n\rangle$.

\begin{figure}
    \centering
    \includegraphics[width=0.45\textwidth]{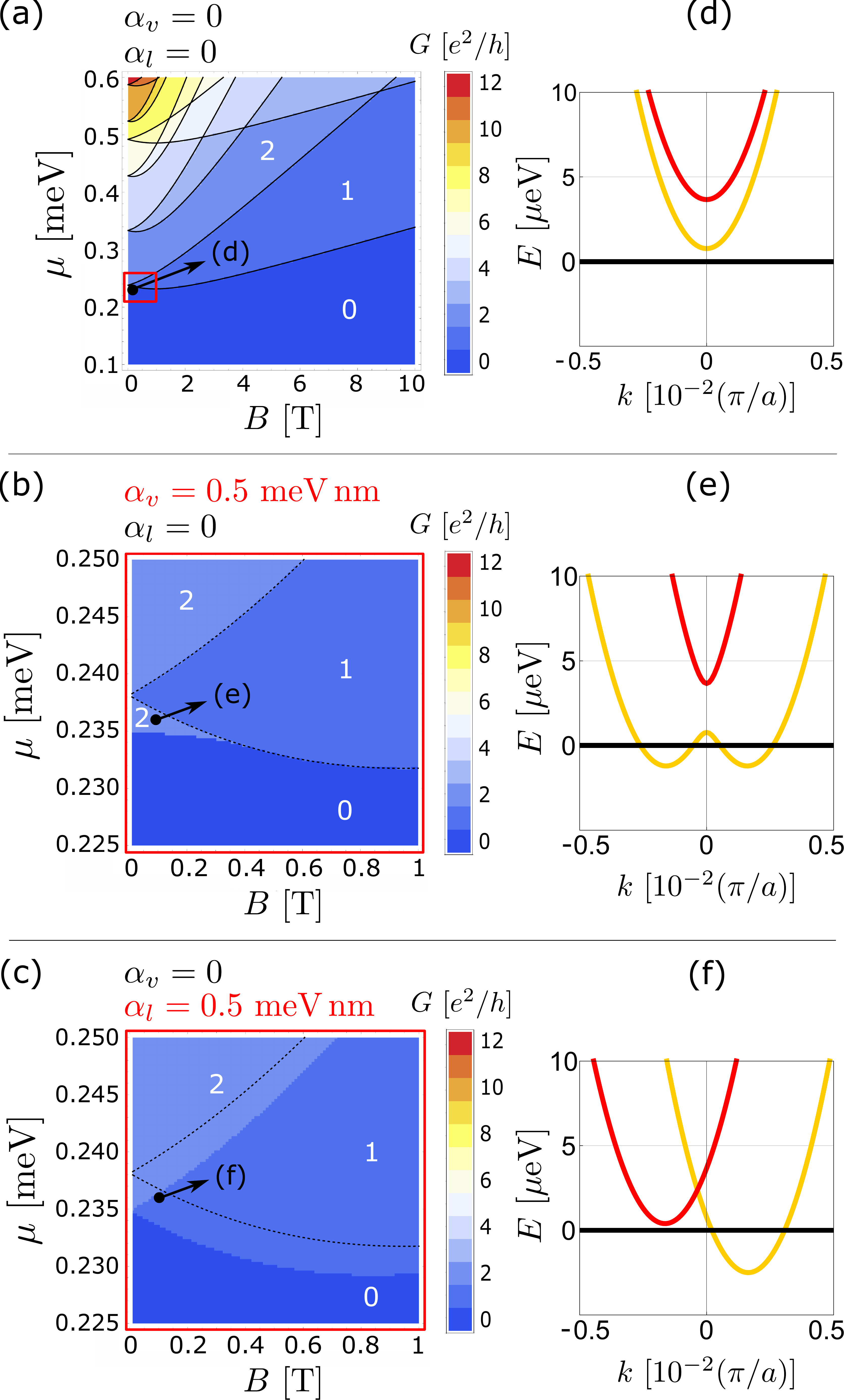}
    \caption{\small{\textbf{Zero-bias conductance with and without SOC.} Conductance $G$ (in units of $e^2/h$) as a function of magnetic field $B$ and chemical potential $\mu$ for $\alpha_v = \alpha_l = 0$ meV$\,$nm (a), $\alpha_v = 0.5$ meV$\,$nm and $\alpha_l = 0$ meV$\,$nm (b), and $\alpha_v = 0$ meV$\,$nm and $\alpha_l = 0.5$ meV$\,$nm (c). The dashed black lines in panels (b) and (c) corresponds to the borders of the conductance regions in the absence of SOC [solid lines in panel (a)]. Panels (d), (e) and (f) show the two dispersion relations associated with the eigenstates of $\langle 0,0,k| H|0,0,k\rangle$ related to panels (a),(b) and (c) respectively for $B = 0.1$ T and $\mu = 0.236$ meV [black dots in panels (a-c)]. We chose typical waveguide parameters~\cite{Annadi2018, Briggeman2020Science} $m_x = m_y = 2 m_e$ where $m_e$ is the electron mass, $l_y = \sqrt{\hbar/(m_y \omega_y)} = 20$ nm, $l_z = \sqrt{\hbar/(m_y \omega_y)} = 10$ nm, $g = 0.5$.}}
    \label{fig3}
\end{figure}

The transport properties of the waveguide can be computed directly from the model above. Of particular interest is the conductance $G(\mu) = \mathrm{d}I(\mu)/\mathrm{d}V$, corresponding to the derivative of the current in the waveguide (which depends on its chemical potential $\mu$) with respect to the bias voltage $V$. We focus here on the conductance at zero-bias ($V= 0$). Increasing the chemical potential increases the conductance by one quantum $e^2/h$ each time a new eigenstate of $H$ is populated, as can be obtained via Landauer theory~\cite{Landauer1957}. In practice, we calculate the eigenvalues of $H$ as a function of $k$ and count the number of times they cross the zero-energy axis, indicating the position of the Fermi momenta. Figure~\ref{fig3} shows the conductance as a function of the out-of-plane magnetic field $B$ and chemical potential $\mu$ of a waveguide without SOC [$\alpha_v = \alpha_l = 0$, panel (a)], with SOC along $\sigma_y$ [$\alpha_v \neq 0$ and $\alpha_l = 0$, panel (b)] and with SOC along $\sigma_z$ [$\alpha_v = 0$ and $\alpha_l\neq 0$, panel (c)], for typical other parameters. In the absence of SOC, the conductance increases monotically as a function of $\mu$ -- in steps of $2e^2/h$ at $B = 0$ and steps of $e^2/h$ for $B > 0$ due to lifting of the spin-degeneracy by the Zeeman term -- as the transverse modes $|m,n\rangle$ are gradually populated. This case has been studied in detail in~\cite{Annadi2018,Briggeman2020Science}. Figure~\ref{fig3}(d) shows the two dispersion relations of the electrons in the transverse mode $|0,0\rangle$ for a value $(B,\mu)$ just below the threshold necessary to populate one of the bands. The presence of SOC perturbs the dispersion relations. For $\alpha_v\neq 0$, a gap opens at low $B$ in the single-particle dispersions relations. This can lead to a situation where the lower one exhibits four crossing with the zero-energy axis [see Fig.~\ref{fig3}(e)], which is known to lead to a conductance of $2e^2/h$ (see e.g.~\cite{Pershin2004}), as shown in Fig.~\ref{fig3}(b). As a consequence, the conductance increase as a function of $\mu$ is no longer monotonic. For $\alpha_l \neq 0$, the dispersion relations become asymmetric in $k$ and are sightly pushed downward. This effect is likely to populate one of the bands as shown in Fig.~\ref{fig3}(f) and thus increase the conductance at lower chemical potential than one would have expected in the absence of SOC.

\subsection{Electron-electron interactions}

We now model the effects of attractive interactions between electrons of opposite spins belonging to different subbands, in order to study how it affects the single-particle picture presented above. Since electrons interact most likely when they have similar single-particle energies, we will focus in the following on regions of parameters where only two subbands labelled as $\alpha = |m,n,\downarrow\rangle$ and $\beta = |m',n',\uparrow\rangle$ are relevant [see e.g.~$|0,0,\uparrow\rangle$ and $|0,0,\downarrow\rangle$ at $B \simeq 0$ in Fig.~\ref{fig3}(a), or $|0,1,\uparrow\rangle$ and $|1,0,\downarrow\rangle$ around $B \simeq 2.5$ T and $\mu \simeq 0.5$ meV]. In second quantization, the interaction Hamiltonian we consider takes the form
\begin{equation}\label{HinteractionSQ}
\begin{aligned}
    H_I &=\sum_{k} \bigg[ \sum_{\gamma = \alpha,\beta} \Sigma_{\gamma} c_{\gamma k}^\dagger c_{\gamma k}  - \left( \chi c_{\alpha k}^\dagger c_{\beta k} + \mathrm{h.c.} \right)\bigg.  \\
   &\quad\quad\quad  \bigg. +  \Delta ( c_{\alpha k}^\dagger c_{\beta  -k}^\dagger - c_{\alpha  k} c_{\beta -k})\bigg],
  \end{aligned}
\end{equation}
where $c_{k \alpha}$ is the annihilation operator of an electron in the subband $\alpha$ with a wavevector $k$, and where $\Sigma_{\alpha}$, $\chi$ and $\Delta$ are the Hartree, Fock and Bogoliubov mean fields defined in a similar fashion than in standard BCS theory~\cite{Tinkham2004} as
\begin{align}\label{mf1}
&\Sigma_{\alpha} = \frac{U}{2\pi} \int_{-\infty}^{\infty} \langle c_{\beta k}^\dagger c_{\beta k}\rangle \,\mathrm{d}k, \\
\label{mf2}
&\chi = \frac{U}{2\pi} \int_{-\infty}^{\infty} \langle c_{\alpha k}^\dagger c_{\beta k}\rangle \,\mathrm{d}k, \\
\label{mf3}
&\Delta = \frac{U}{2\pi} \int_{-\infty}^{\infty}  \langle c_{\alpha k} c_{\beta -k} \rangle \,\mathrm{d}k,
\end{align}
which have to be found self-consistently as explained below, with $U$ the interaction strength. In our model, we consider that $U$ has the following empirical scaling with the magnetic field 
\begin{equation}\label{Uscaling}
U \equiv U(B) = U_0 \sqrt{1-\frac{\omega_c^2}{\Omega^2}} = U_0 \frac{\omega_y}{\Omega},
\end{equation}
where $U_0$ is a bare interaction strength (in dimensions of energy $\times$ length). This makes $|U|$ decreasing as a function of the magnetic field. A physical justification is the following: the effective kinetic energy of the electrons of our model [see Eq.~(\ref{Emnk})] scales with the magnetic field as $\omega_y^2/\Omega^2$, so that electron momenta behave as $k \sim \sqrt{2 m_x E/\hslash^2}(\Omega/\omega_y)$. The scaling of $U$ is such that it compensates the increase with magnetic field of the typical range of momenta involved in the interactions. This has the effect of keeping the mean-fields~(\ref{mf1})-(\ref{mf3}) $\Delta, \Sigma,\chi \propto U \int \cdot \, dk$ independent of this effective scaling (it amounts to a contraction of the real-space quantization length $L$ reciprocally associated with $k$). Our phenomenological scaling has allowed us to obtain results in qualitative agreement with experimental data~\cite{briggeman2019vertical}. Note that while other scalings with the magnetic field could be chosen and possibly compatible with experimental data, a parameter $U$ without scaling with $B$ at all would lead to the unphysical situation where the resulting pairing of electrons $\Delta$ would increase with $B$ in an unbounded way. Note also that the findings presented below remain qualitatively the same for other choices of scaling. Finally, in the case of SOC produced via modulations, it would be reasonable to consider a modulated interaction as in~\cite{Shavit2020Modulation}, where it has been shown that such assumption can produce fractional conductance plateaux, compatible with the experimental data presented in~\cite{briggeman2019lateral,briggeman2019vertical}.

Note that an interaction Hamiltonian of the form~(\ref{HinteractionSQ}) can be viewed as originating from a contact interaction, which is a  standard description of interactions in dilute cold atomic gases. In Appendix B, we present a derivation of Eq.~(\ref{HinteractionSQ}) from such an assumption, relevant for these systems.

\subsection{Hartree-Fock-Bogoliubov Model}

We now add to Eq.~(\ref{HinteractionSQ}) the single-particle Hamiltonian~(\ref{Hwaveguide}) for only two subbands $\alpha$ and $\beta$. In second quantization, this yields the Hartree-Fock-Bogoliubov Hamiltonian
\begin{equation}\label{HtotSQ}
\begin{aligned}
    &H_{HFB} = H + H_I \\
    &=\sum_{k} \bigg[ \sum_{\gamma = \alpha,\beta} \Big(\xi_{\gamma k} + \Sigma_{\gamma} + 2 s(\gamma) \frac{\omega_y^2}{\Omega^2} \alpha_l k \Big) c_{\gamma k}^\dagger c_{\gamma  k} \bigg. \\ 
    &\quad\quad\quad+ \Big[ \Big(i \alpha_v \frac{\omega_y^2}{\Omega^2} k  - \chi \Big)c_{\alpha k}^\dagger c_{\beta k}  + \mathrm{h.c.} \Big]  \\
   &\quad\quad\quad  \bigg. + \Delta \Big( c_{\alpha  k}^\dagger c_{\beta  -k}^\dagger - c_{\alpha  k} c_{\beta  -k}\Big)\bigg].
  \end{aligned}
\end{equation}
As can be seen in Eq.~(\ref{HtotSQ}), the Hartree shifts and the SOC coming from a lateral electric field alter the single-particle energies $\xi_{\alpha k}$. More specifically, the Hartree shifts $\Sigma_{\alpha}$ push down the energy of the electron in the subband $\alpha$ due to the presence of the other electron in the subband $\beta$. Regarding the Fock fields, they alter the same spin-flip terms as the SOC coming from a vertical electric field. 

The Hamiltonian $H_{HFB}$ can be written in the electron and hole basis $\{ c_{\alpha k},c_{\alpha -k}^\dagger,c_{\beta k},c_{\beta -k}^\dagger\}$ and defines the following self-consistent eigenvalue problem
\begin{widetext}
\begin{equation}\label{HSQMF}
    \begin{pmatrix}\xi_{\alpha k} + \Sigma_{\alpha} +  \alpha_l \frac{\omega_y^2}{\Omega^2} k  &0&-i \alpha_v \frac{\omega_y^2}{\Omega^2} k - \chi & \Delta \\
    0 & -\xi_{\alpha k}-\Sigma_{\alpha}+ \alpha_l \frac{\omega_y^2}{\Omega^2} k &-\Delta & i \alpha_v \frac{\omega_y^2}{\Omega^2} k - \chi \\
    i\alpha_v \frac{\omega_y^2}{\Omega^2} k -\chi &-\Delta & \xi_{\beta k} + \Sigma_{\beta} -  \alpha_l \frac{\omega_y^2}{\Omega^2} k &0\\
    \Delta & -i\alpha_v \frac{\omega_y^2}{\Omega^2} k-\chi & 0 &-\xi_{\beta k}-\Sigma_{\beta} - \alpha_l \frac{\omega_y^2}{\Omega^2} k
    \end{pmatrix} \phi_{j k}= E_{j k} \phi_{j k},
\end{equation}
\end{widetext}
where $E_{j k}$ and $\phi_{j k}$ ($j = 1,2$) are the quasi-energies and the quasi-particle wavefunctions (note that there are also two other solutions of the eigenvalue problem corresponding to the associated quasi-holes). Equation~(\ref{HSQMF}) is the direct generalization of Eq.~(5) of \cite{Annadi2018} to include SOC.
In order to solve Eq.~(\ref{HSQMF}), we need to proceed self-consistently, by starting with an initial guess of the values of the mean fields $\Sigma_{\alpha}$, $\chi$ and $\Delta$. These values are inserted into Eq.~(\ref{HSQMF}), which is then solved to obtain the quasi-energies and quasi-particle wavefunctions $E_{j k}$ and $\phi_{j k}$ and compute new values for the mean fields using Eqs.~(\ref{mf1})-(\ref{mf3}). This last step is done via Bogoliubov transformation of the electrons annihilation operators $c_{\alpha k}$ and $c_{\beta k}$ into quasi-particle annihilation operators $\gamma_{1 k}$ and $\gamma_{2 k}$, where we assume here the thermal state correlation functions
\begin{equation}
\begin{aligned}
&\langle \gamma_{i k}^\dagger \gamma_{j k} \rangle = \delta_{ij} n(E_{ik}),\\
&\langle \gamma_{i k} \gamma_{j k}^\dagger \rangle = \delta_{ij} [1- n(E_{ik})],\\
&\langle \gamma_{i k} \gamma_{j k} \rangle = \langle \gamma_{i k}^\dagger \gamma_{j k}^\dagger \rangle = 0,\\
\end{aligned}
\end{equation}
where $n(E) = 1/[1+ e^{E/(k_B T)}]$ is the Fermi distribution with $k_B$ the Boltzmann constant and $T$ the temperature. The procedure is then repeated until the mean fields have converged. Note that since we work in the electron-hole basis (twice as big as the physical basis), the computed quasi-energies always appear in conjugate pairs ($E_{1k}$, $-E_{1-k}$) and ($E_{2k}$, $-E_{2-k}$), and one has to select only one member of each pair. 

\section{Results}

\subsection{Without SOC}

\begin{figure*}
    \centering
    \includegraphics[width=\textwidth]{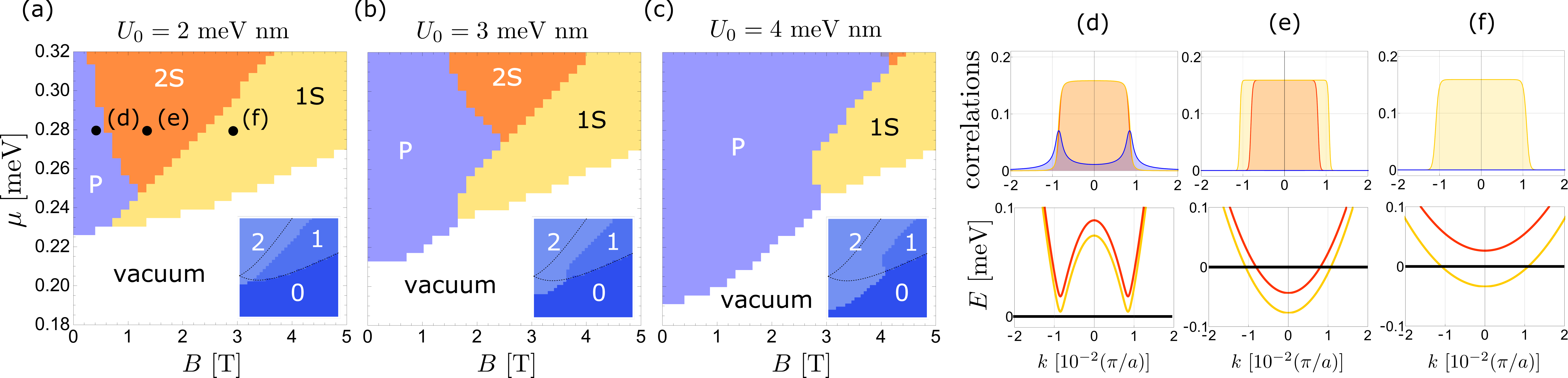}
    \caption{\small{\textbf{Phase diagrams and zero-bias conductance with interactions and without SOC.} (a-c), Phase diagrams of the waveguide near the crossing of the subbands $\alpha = |0,0,\downarrow\rangle$ and $\beta =|0,0,\uparrow\rangle$ as a function of magnetic field $B$ and chemical potential $\mu$ for different attractive interaction strengh $|U_0| = 2$ meV$\,$nm (a), $3$ meV$\,$nm (b), and $4$ meV$\,$nm (c), and temperature $T = 25$ mK. The phase `P' corresponds to a region where $\Delta > 10^{-3}$ meV ($ \sim k_B T$), i.e., when we have electron pairs. The phase `1S' and `2S' denote the phases where one and two single-particle bands are populated, respectively. The insets show the associated conductance (in units of $e^2/h$), where the thin dashed lines correspond to the non-interacting case. Panels (d-f) show on the top the correlations $\langle c_{\beta k}^\dagger c_{\beta k}\rangle$ (yellow), $\langle c_{\alpha k}^\dagger c_{\alpha k}\rangle$ (red) and $\langle c_{\alpha k} c_{\beta -k} \rangle$ (blue) appearing in the definitions of the mean-fields~(\ref{mf1})-(\ref{mf3}). The corresponding band structures appear at the bottom, showing e.g. in panel (d) (bottom) that a gap $\Delta$ protects the band from single-particle excitations. Other parameters as in Fig.~\ref{fig3}.}}
    \label{fig4}
\end{figure*}

We first show the effects of interactions in the absence of spin-orbit coupling ($\alpha_v = \alpha_l =0$). We focus here on the two lowest subbands $\alpha = |0,0,\downarrow\rangle$ and $\beta = |0,0,\uparrow\rangle$ that are close from each other at low magnetic field. We solve the eigenvalue problem~(\ref{HSQMF}) as a function of $B$ and $\mu$ for different interaction strength $U_0$ until the single-particle spectra and the mean-fields have converged, which allows us to construct phase diagrams and associated conductance maps, as shown in Figure~\ref{fig4}. As for Figure~\ref{fig3}, the available single-particle phases are identified from the number of crossing points with the zero-energy axis (i.e. the number of finite Fermi momenta) of the converged single-particle spectra, which indicates the presence or not of electrons in the single-particle bands. A conductance of $e^2/h$ is associated with each occupied band. By contrast, a pair phase is identified when a non-zero value of $\Delta$ emerges from the calculations, and is associated with a conductance of $2e^2/h$. This corresponds to the situation where the single-particle spectra are gapped. Figure~\ref{fig4}(a-c) shows the obtained phase diagrams and conductance maps (insets) for different $U_0$. The insets show the associated conductance, where the conductance associated with a pair is indistinguishable from the conductance associated with the two unpaired electrons in different subbands. Due to the presence of the interactions, the electrons in the two first subbands are paired at low $B$, leading to a direct increase of $2e^2/h$. For a certain value of $B$ defined as the pairing field $B_p$, the electron pairs are finally split, leaving us with signatures of single-electron subbands. Increasing $|U_0|$ enhanced the pairing area and thus shifts the position where the conductance lines splits. In our model, the resulting pairing energy behaves as $\Delta \propto U^2(0) m_x$, in agreement with the expected behaviour $\Delta \propto U^2(0)/t$ of a tight-binding model with hopping parameter $t = \hslash^2/(m_x a^2)$ and lattice spacing $a$~\cite{Briggeman2020Science}. The pairs are broken when the Zeeman splitting energy $g \mu_B B$ compensates the pairing energy $\Delta$, which provides the typical scaling of the pairing field $B_p \propto U^2(0) m_x^*/(g \mu_B)$ at which the lines split.

Figure~\ref{fig4}(d-f) shows the values of the correlations functions $\langle c_{\beta k}^\dagger c_{\beta k}\rangle$, $\langle c_{\alpha k}^\dagger c_{\alpha k}\rangle$ and $\langle c_{\alpha k} c_{\beta -k} \rangle$ appearing in the mean-fields $\Sigma_{\alpha}$, $\Sigma_{\beta}$, and $\Delta$ for three pairs of values ($B$,$\mu$) corresponding to three different points of the phase diagram in Fig.~\ref{fig4}(a) along the horizontal line corresponding to $\mu = 0.28$ meV, one with pairing ($B = 0.5$ T), one without pairing but the two subbands populated ($B = 1.5$ T), and one without pairing and only one subband populated ($B = 3$ T). In the absence of pairing, the correlations functions appearing in the Hartree shifts correspond to standard single-particle Fermi distributions and the one appearing in the definition of the pairing vanishes [see Fig.~\ref{fig4}(e) and (f)]. In the presence of pairing, the single-particle distributions are smoothen around the Fermi level, where non-vanishing pairing correlations appear [see Fig.~\ref{fig4}(d)], in a similar way than in standard BCS theory~\cite{Tinkham2004, de_gennes_superconductivity_2018}. The associated single-particle spectra plotted at the bottom of the panels (d-f) show clearly when electrons occupy one of the single-particle band (f),  the two single-particle bands (e), or when they are paired, as translated by the apparition of a gap that protects the bands from single-particle excitations (d). 

Note that the transport properties of the waveguide are strongly dependent on the temperature. Typical experiments in LAO/STO devices work with temperature of a few dozen of milli-Kelvin, and we chose $T = 25$ mK throughout our work. Increasing the temperature destroys electron pairs typically when the thermal fluctuations overcome the pairing energy, i.e., when $k_BT \gtrsim \Delta$.

\subsection{With SOC}

\subsubsection{Vertically-induced SOC ($\alpha_v \neq 0$)}
\begin{figure*}
    \centering
    \includegraphics[width=\textwidth]{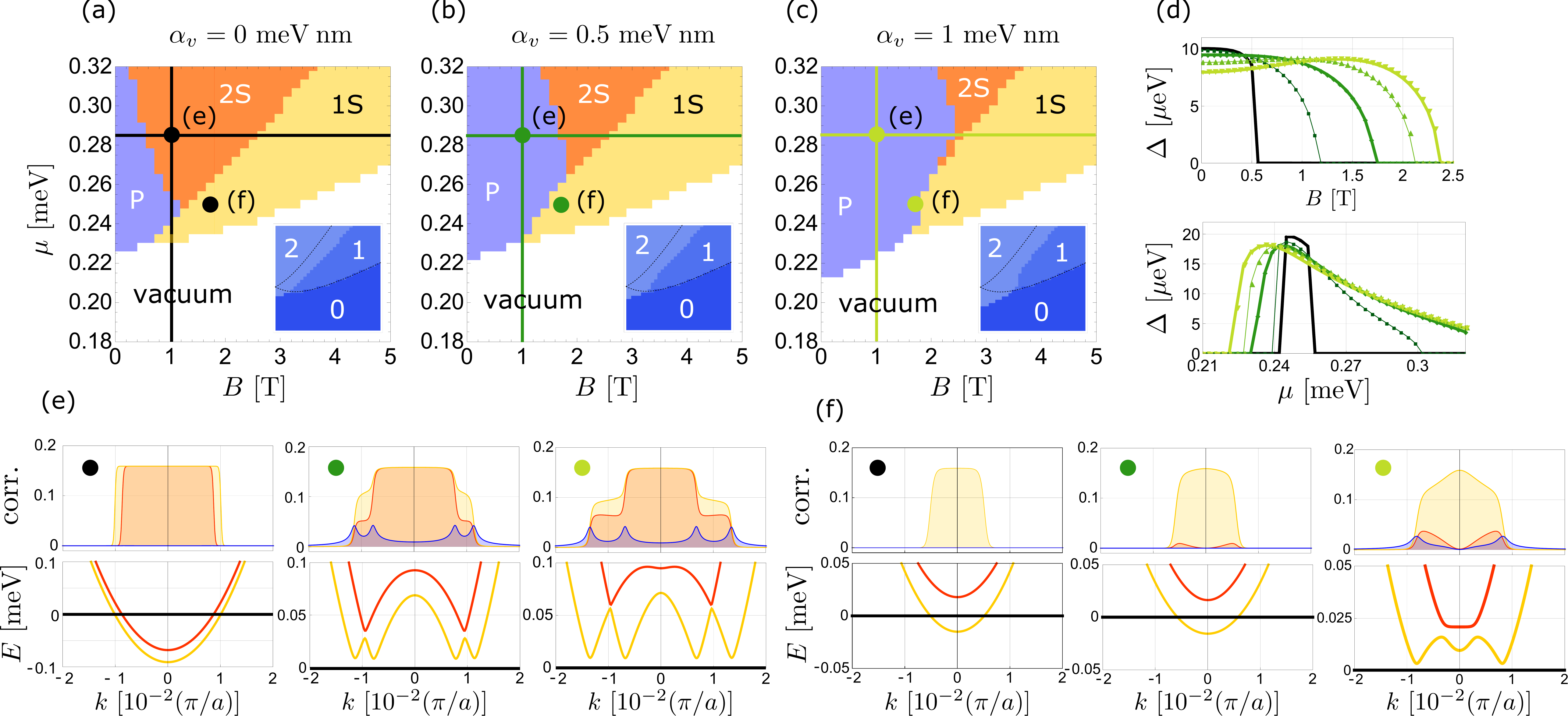}
    \caption{\small{\textbf{Phase diagrams and zero-bias conductance with interactions and vertically-induced SOC ($\alpha_v \neq 0$).} (a-c) Phase diagrams of the waveguide near the crossing of the subbands $\alpha = |0,0,\downarrow\rangle$ and $\beta =|0,0,\uparrow\rangle$ as a function of magnetic field $B$ and chemical potential $\mu$ for different SOC strengh $\alpha_v = 0$ meV$\,$nm (a), $0.5$ meV$\,$nm (b), and $1$ meV$\,$nm (c), for interaction strength $|U_0| = 2$ meV$\,$nm and temperature $T = 25$ mK. The phases are labelled as in Fig.~\ref{fig4}, and the insets show the associated conductance. (d) Pairing $\Delta$ across horizontal (top) and vertical (bottom) linecuts of the phase diagrams for SOC strength ($\alpha_v = 0$, $0.25$, $0.5$, $0.75$ and $1$ meV$\,$nm). (e),(f) (top) Correlations $\langle c_{\beta k}^\dagger c_{\beta k}\rangle$ (yellow), $\langle c_{\alpha k}^\dagger c_{\alpha k}\rangle$ (red) and $\langle c_{\alpha k} c_{\beta -k} \rangle$ (blue) appearing in the definitions of the mean-fields~(\ref{mf1})-(\ref{mf3}). (e),(f) (bottom) Corresponding band structures. Other paramaters as in Fig.~\ref{fig3}.}}
    \label{fig5}
\end{figure*}

We now study the interplay between SOC and interactions in such a waveguide model. We start with the case of a SOC engineered through an applied vertical electric field ($\alpha_v \neq 0$). Figure~\ref{fig5}(a-c) shows phase diagrams and associated conductance maps (in units of $e^2/h$) for different values of $\alpha_v$ and for $|U_0| = 2$ meV$\,$nm. As can be seen in the figures, an enhanced pairing area is obtained for increasing $\alpha_v$, showing that the SOC can assists pairing. This is clearly seen in panel (d) showing $\Delta$ across horizontal (top) and vertical (bottom) line cuts of the previous panels. Intuitively, this can be understood since the SOC term is proportional to $\sigma_y$, it provides a direct coupling between up and down electrons. In order to understand this in more detail, we plot in \ref{fig5}(e) and (f) the correlations functions $\langle c_{\beta k}^\dagger c_{\beta k}\rangle$, $\langle c_{\alpha k}^\dagger c_{\alpha k}\rangle$ and $\langle c_{\alpha k} c_{\beta -k} \rangle$ appearing in the definitions of the mean-field $\Sigma_{\alpha}$, $\Sigma_{\beta}$, and $\Delta$ for two pairs of values $(B,\mu)$ corresponding to two specific points in the phase diagrams: ($B = 1$ T, $\mu = 0.285$ meV) [panels in (e)] and ($B = 1.75$ T, $\mu = 0.24$ meV) [panels in (f)]. The different panels in (e) and (f) show respectively the transition occuring from the phases `2S' to `P' (e) and `1S' to `P' (f) when increasing the SOC strength. More specifically, we see that the single-particle subband is linearly depleted in favor of the second single-particle subband, even if the latter is non-occupied without SOC. The physical picture is thus that the SOC is able to transfer electrons of one subband to the other one [see e.g. middle plot in panel (f)), which are then likely to pair due to the interactions [see e.g. right plot in panel (f)]. As mentioned in the introduction, this phenomenon is a possible interpretation of the experimental data presented in~\cite{briggeman2019vertical}, showing that vertical modulations of the c-AFM tip producing the electron waveguide lead to an enhanced pairing of the two lowest subbands up to very large magnetic field (as large as $B = 16$ T in some devices) compared to unmodulated devices. More broadly, our findings are consistent with other studies, as enhanced pairing due to SOC has been found in other contexts, such as in superconductors with magnetic impurities~\cite{Wei2006Enhancing} and in cold atoms~\cite{DellAnna2011,DellAnna2012}.

\subsubsection{Laterally-induced SOC ($\alpha_l \neq 0$)}

\begin{figure*}
    \centering
    \includegraphics[width=\textwidth]{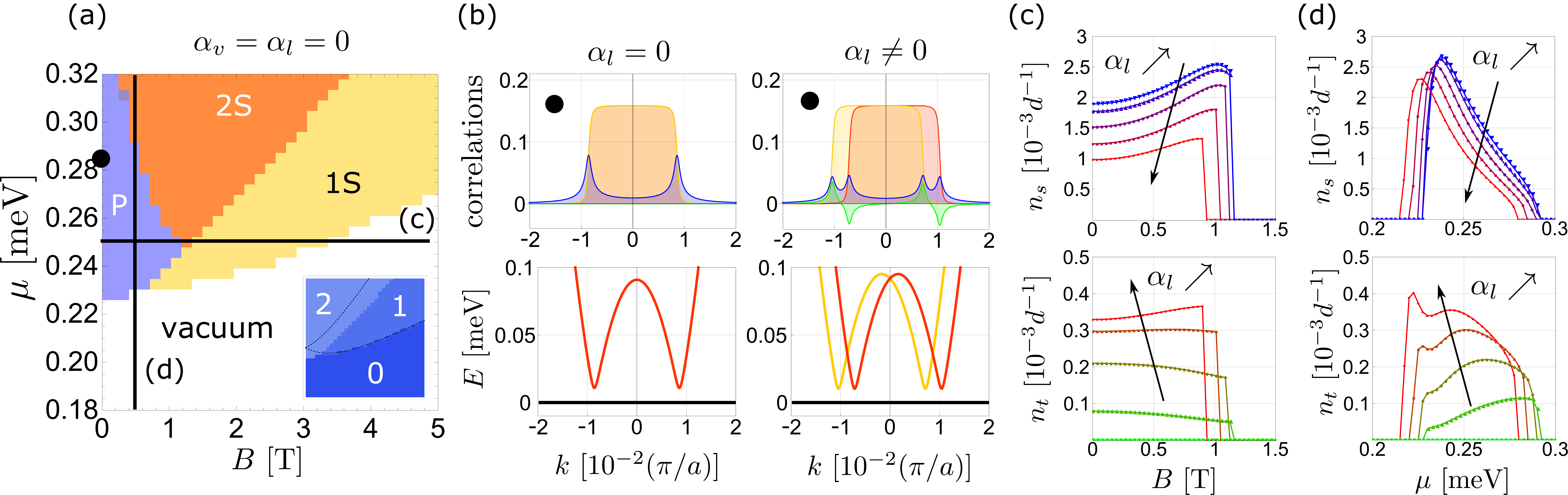}
    \caption{\small{\textbf{Emergence of spin-triplet electron pairs via laterally-induced SOC ($\alpha_v \neq 0$).} (a) Phase diagram of the waveguide near the crossing of the subbands $\alpha = |0,0,\downarrow\rangle$ and $\beta =|0,0,\uparrow\rangle$ as a function of magnetic field $B$ and chemical potential $\mu$ for SOC strengh $\alpha_v = \alpha_l = 0$ meV$\,$nm, interaction strength $|U_0| = 2$ meV$\,$nm and temperature $T = 25$ mK [same as~\ref{fig4}(a) and \ref{fig5}(a)]. (b) Correlations $\langle c_{\beta k}^\dagger c_{\beta k}\rangle$ (yellow), $\langle c_{\alpha k}^\dagger c_{\alpha k}\rangle$ (red), $\langle c_{\alpha k} c_{\beta -k} - c_{\beta k} c_{\alpha -k}\rangle/2$ (blue) and $\langle c_{\alpha k} c_{\beta -k} + c_{\beta k} c_{\alpha -k}\rangle/2$ (green) as a function of $k$ for $B = 0$ and $\mu = 0.285$ meV$\,$nm [black dot in panel (a)] and SOC strength $\alpha_l = 0$ (left) and $\alpha_l = 0.5$ meV$\,$nm (right). The SOC induces the generation of spin-triplet electron pairs. (c) Density of singlet (top) and triplet (bottom) as a function of $B$ for $\mu= 0.25$ meV$\,$nm along the horizontal black line of panel a for $\alpha_l = 0$, $0.25$, $0.5$, $0.75$, $1$ meV$\,$nm [from blue to red (top) and from green to red (bottom)]. (d) Density of spin-singlet (top) and spin-triplet (bottom) as a function of $\mu$ for $B = 0.5$ T along the vertical black line of panel a for $\alpha_l = 0$, $0.25$, $0.5$, $0.75$, $1$ meV$\,$nm [from blue to red (top) and from green to red (bottom)]. Other parameters as in Fig.~\ref{fig3}.}}
    \label{fig6}
\end{figure*}

We now investigate the case of SOC coming from an applied electric field along $y$ ($\alpha_l \neq 0$). In contrast to a vertically induced SOC, a laterally-induced SOC does not lead to an increase of the pairing area `P'. However, it allows for the emergence of triplet pairs of electrons in the area where the usual pairing defined through non-zero values of $\Delta$ is usually observed, which is a singlet pairing by construction. Indeed, using the fermionic nature of the electron annihilation operators, we can rewrite without restriction
\begin{equation}
\begin{aligned}
\Delta &= \frac{U}{2\pi} \int  \langle c_{\alpha k} c_{\beta -k} \rangle \,\mathrm{d}k \\
&= \frac{U}{4\pi}\int  \left[ \langle c_{\alpha k} c_{\beta -k} \rangle - \langle c_{\beta -k} c_{\alpha k} \rangle\right]  \mathrm{d}k,\\
&= \frac{U}{4\pi}\int  \left[ \langle c_{\alpha k} c_{\beta -k} \rangle - \langle c_{\beta k} c_{\alpha -k} \rangle\right]  \mathrm{d}k,\\
\end{aligned}
\end{equation}
showing that $\Delta$ comes from the expectation value of the spin-singlet pair operator 
\begin{equation}\label{singletO}
s_k = \frac{c_{\alpha k} c_{\beta -k} - c_{\beta k} c_{\alpha -k}}{\sqrt{2}}.
\end{equation} 
Due to the presence of a laterally-induced SOC, it is legitimate (see e.g.~\cite{DellAnna2011,DellAnna2012}) to also look at the presence of spin triplet pairs defined through the spin-triplet operator
\begin{equation}\label{tripletO}
t_k = \frac{c_{\alpha k} c_{\beta -k} + c_{\beta k} c_{\alpha -k}}{\sqrt{2}}.
\end{equation}
In the absence of a laterally-induced SOC, the single-particle spectra are symmetric under the transformation $k \to -k$ [see e.g., Fig.~\ref{fig3}(d) and (e)] and we thus have $\langle c_{\beta k} c_{\alpha -k} \rangle = \langle c_{\beta -k} c_{\alpha k} \rangle = - \langle c_{\alpha k} c_{\beta -k} \rangle$ so that $\langle t_k \rangle = 0 \, \forall k$. In other words, there is no spin-triplet pairs. However, for $\alpha_l \neq 0$, the single-particle spectra become asymmetric [see e.g., Fig.~\ref{fig3}(f)] so that $\langle c_{\beta k} c_{\alpha -k} \rangle \neq \langle c_{\beta -k} c_{\alpha k} \rangle$ and non-zero values of $\langle t_k \rangle$ are likely to emerge, signalling the presence of triplet pairs in the waveguide. This is what we observed in Fig.~\ref{fig6}. The panel (b) shows the correlations $\langle c_{\beta k}^\dagger c_{\beta k}\rangle$, $\langle c_{\alpha k}^\dagger c_{\alpha k}\rangle$, $\langle c_{\alpha k} c_{\beta -k} - c_{\beta k} c_{\alpha -k}\rangle/2$ and $\langle c_{\alpha k} c_{\beta -k} + c_{\beta k} c_{\alpha -k}\rangle/2$ as a function of $k$ without SOC (left) and with SOC (right), for $|U_0 = 2$ meV$\,$nm and $B = 0$. Note that the phase diagram in panel (a) is the same as in Fig.~\ref{fig4}(a) and \ref{fig5}(a) and is there for comparison. The correlations in panel (b) are evaluated at the coordinates $B$ and $\mu$ of the black dot in panel (a). The SOC splits the Fermi distributions of the single particles and the triplet pair (green) that appears is an odd function of $k$, by contrast with the singlet pair (blue) which is even, as expected (see e.g.~\cite{Sigrist1991}). As a consequence, any pairing term of a form $~\propto \int \mathrm{d}k \langle t_k \rangle $ (similar to the singlet pairing $\Delta$) vanishes. We compare here the densities of singlet and triplet pairs defined as
\begin{equation}
\begin{aligned}
n_s = \frac{1}{2\pi}\int \left|\langle s_k\rangle\right|^2 
 \mathrm{d}k, \\
n_t = \frac{1}{2\pi}\int \left|\langle t_k\rangle\right|^2
 \mathrm{d}k,
\end{aligned}
\end{equation}
in Fig.~\ref{fig6} as a function of $B$ for $\mu = 0.25$ meV$\,$nm (c) and as a function of $\mu$ for $B = 0.5$ T (d), for $|U_0| = 2$ meV$\,$nm and different SOC strength $\alpha_l$. The SOC can generate a non-negligible fraction of triplet pairs compared to singlet pairs. Also, it shifts the pairing area slightly downward along $\mu$, and contracts it as a function of $B$. Hence, it does not increase the critical magnetic fields defining the pairing area. This is consistent with experimental observations in laterally-modulated waveguides that are believed to generate such a SOC~\cite{briggeman2019lateral}, since such devices do not exhibit enhanced pairing fields compared to unmodulated waveguide. Our model suggests however that triplet pairing could potentially be observed at low magnetic field and signatures of this could thus be investigated in future experiments.

\section{Conclusion}

We showed that the confinement of the electrons in 1D channels (as is found in waveguides produced e.g. at an LAO/STO interface) is likely to reduce the intrinsic Rashba SOC felt by the electrons in the lowest band. Then, after discussing ways to recover and engineer different forms of SOC, we showed that spin-singlet and spin-triplet electron pairs can be controlled via SOC mechanisms. In particular, we showed that vertically-induced SOC, which are of the Rashba form, can stabilize spin-singlet electron pairs over a larger range of applied magnetic field and gate-tunable chemical potential. By contrast, laterally-induced SOC does not increase the parameter space where spin-singlet electron pairs are stable --- it even contracts it --- but rather, generates spin-triplet electron pairs. Our model is based on a self-consistent Hartree-Fock-Bogoliubov that we generalised to include SOC, and our results are consistent with recent experiments~\cite{briggeman2019lateral,briggeman2019vertical} that are believed to engineered the forms of SOCs investigated in this work. 
 
Our work provides a consistent theoretical framework to study pairing mechanisms in 1D waveguides with attractive interactions and SOC. It would, however, be interesting in the future to explore different models for the interactions, or go beyond mean-field to study situations with stronger electron-electron interactions. Also, it would be interesting to explore the coupling between different transverse modes that appear for larger SOC strength. 

Our results showed that the collective spin of electrons pairs can be controlled via SOC. Knowing that bound states of more than two electrons can be realised in LAO/STO devices~\cite{Briggeman2020Science}, it would be worthwhile to explore how SOC could produce and engineer exotic collective spin states of higher spin quantum numbers. In addition, with the simultaneous presence of SOC, interactions and magnetic field, 1D waveguides written on the LAO/STO interface constitute interesting candidates for investigating the physics of Majorana fermions~\cite{Leijnse2012, Mazziotti2018Majorana}.

Finally, our findings could be explored in other 1D platforms for fermionic particles with attractive interactions, such as cold atoms. This could make it possible to explore different forms of interactions, and provide new ways to control the phases and transport properties in these systems using e.g., dissipation engineering~\cite{DamanetNJP2019,DamanetPRL2019}, measurements~\cite{Laflamme2017Continuous, Uchino2018Universal} and feedback~\cite{Muldoon2012Control, Morrow2002Feedback}. The diagnostic of the impact of dissipation on these systems constitutes thus an interesting future project and a first step in these perspectives.

\begin{acknowledgements}
Work at the University of Strathclyde was supported by
the EPSRC Programme Grant DesOEQ (EP/P009565/1),
and by AFOSR Grant No. FA9550-18-1-
0064. J.\ Levy acknowledges support from a Vannevar Bush Faculty Fellowship (ONR N00014-15-1-2847), and the National Science Foundation (PHY-1913034).
\end{acknowledgements}

\section*{Appendix A}

We discuss here in which conditions the quadratic terms in $\alpha_v$ and $\alpha_l$ in Eqs.~(\ref{Hmotional}) and the last term of~(\ref{COM}) can be neglected.

\begin{figure}
    \centering
    \includegraphics[width=0.35\textwidth]{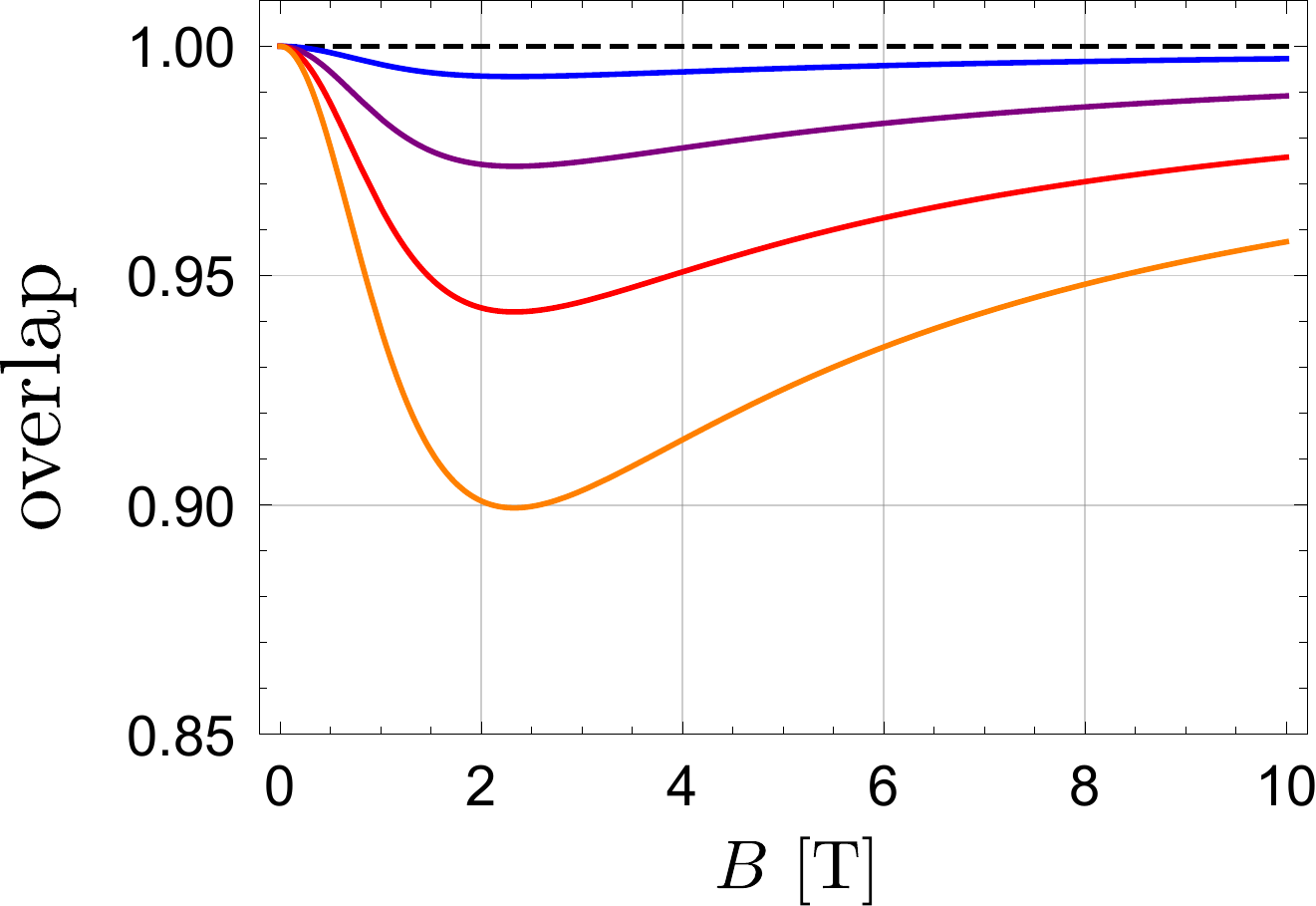}
    \caption{\small{Overlap [Eq.~(\ref{alphaB})] between the two spin-dependent eigenstates of along $y$ as a function of magnetic field $B$ for $\alpha_l = 0$ (dashed black), $0.25$ (blue), $0.5$ (purple), $0.75$ (red) and $1$ (orange) meV$\,$nm, for $m_y = 2 m_e$, where $m_e$ is the electron mass and $l_y = \sqrt{\hbar/(m_y \omega_y)} = 20$ nm.}}
    \label{fig:app1}
\end{figure}

Suppose that $\alpha_v = 0$ and $\alpha_l \neq 0$. The eigenstate of the confining potential along $y$ will be of the form
\begin{multline}
\phi_m^\sigma(y) = \left( \frac{\Omega m_y}{\pi\hbar}\right)^{\frac{1}{4}} \frac{1}{2^m m!} e^{-\frac{m_y \Omega}{2\hbar}(y-y_0^\sigma(k))^2} \\ H_m\left( \sqrt{\frac{m_y\Omega}{\hbar}} [y - y_0^\sigma(k)]\right)  
\end{multline}
where $H_m$ are the standard Hermite polynomials and where $y_0^\sigma(k)$ ($\sigma = \downarrow,\uparrow$) are their spin-dependent center-of-masses
\begin{equation}
\begin{aligned}
y_0^\uparrow(k)  &\equiv \langle \uparrow | y_0(k) | \uparrow \rangle = \frac{\hbar \omega_c k}{\sqrt{m_y m_x} \Omega^2} + \frac{e B}{\hbar m_y \Omega^2} \alpha_l, \\
y_0^\downarrow(k)  &\equiv \langle \downarrow | y_0(k) | \downarrow \rangle = \frac{\hbar \omega_c k}{\sqrt{m_y m_x} \Omega^2} - \frac{e B}{\hbar m_y \Omega^2} \alpha_l. \\
\end{aligned}
\end{equation}
We study below in which conditions the two eigenstates will have an overlap close to the identity, which will correspond to the parameter regime where their spin-dependence can be neglected. For the ground state ($m = 0$), we have
\begin{equation}\label{alphaB}
\int_{-\infty}^{\infty} \phi_0^\uparrow(y) \phi_0^{\downarrow*}(y)dy  = e^{-\frac{m_y(y_0^\uparrow(k) - y_0^\downarrow(k))^2 \Omega}{4\hbar}}=  e^{-\frac{e^2 B^2 \alpha_l^2}{m_y \hbar^3\Omega^3}} 
\end{equation}
Figure~\ref{fig:app1} shows the right-hand-side of Eq.~(\ref{alphaB}) as a function of the magnetic field $B$ for different values of $\alpha_l$ and for $m_y = 2 m_e$, where $m_e$ is the electron mass and $l_y = \sqrt{\hbar/(m_y \omega_y)} = 20$ nm. As can be seen in the figure, the overlap factor is equal to one within $10\%$ error maximum for $\alpha_l \lesssim 1$ meV$\,$nm. It is thus reasonable to neglect the spin-dependent term of the center-of-mass positions for small spin-orbit coupling strength $\alpha_l$. For consistency, we also neglect the term $-e^2 B^2  \alpha_l^2/(2 m_y \Omega^2\hbar^2)$ in Eq.~(\ref{Hmotional}), which is small in a similar way compared to the energy $\hbar\Omega$ of the effective harmonic trapping along $y$, as we have $e^2 B^2 \alpha_l^2/(2 m_y \hbar^3\Omega^3) \ll 1$ [compare to Eq.~(\ref{alphaB})]. The reasoning above also holds for $\alpha_v \neq 0$ and $\alpha_l = 0$, as it yields the same order of magnitude.

\section*{Appendix B}

We derive here the interaction Hamiltonian~(\ref{HinteractionSQ}) from the assumption of contact interactions, as it is usually the case in cold atoms.

The starting point is the interaction Hamiltonian\begin{equation}\label{Hinteraction}
    H_I= V \int \Psi_{\alpha}^\dagger(x)\Psi_{\beta}^\dagger(x)\Psi_{\beta}(x)\Psi_{\alpha}(x) \mathrm{d}x,
\end{equation}
where $V < 0$ is an attractive interaction strength and $\Psi_{\alpha}(x)$ and $\Psi_{\beta}(x)$ are the field operators for electrons of spin $\sigma$ in the subbands $\alpha$ and $\beta$, respectively. 

As a first approximation, we perform a mean field treatment similar to the one used in BCS theory that supposes that some operator $O$ can be written as $O=\langle O\rangle + \delta O$, i.e., quantum fluctuations $\delta O$ around an expectation value $\langle O\rangle$. By performing this treatment on pairs of operators, we obtain a quadratic form for the interactions. Explicitly, using Wick's theorem on Eq.~(\ref{Hinteraction}), we have
\begin{equation}\label{ansatz}
    \begin{aligned}
    \Psi_{\alpha}^\dagger\Psi_{\beta}^\dagger\Psi_{\beta}\Psi_{\alpha} \approx & \langle \Psi_{\alpha}^\dagger\Psi_{\beta}^\dagger \rangle \Psi_{\beta}\Psi_{\alpha} 
    +\langle \Psi_{\beta}\Psi_{\alpha}\rangle \Psi_{\alpha}^\dagger\Psi_{\beta}^\dagger\\ - & \langle \Psi_{\alpha}^\dagger\Psi_{\beta}\rangle \Psi_{\beta}^\dagger\Psi_{\alpha} - \langle \Psi_{\beta}^\dagger\Psi_{\alpha}\rangle \Psi_{\alpha}^\dagger \Psi_{\beta} \\
    + &\langle \Psi_{\alpha}^\dagger \Psi_{\alpha}\rangle \Psi_{\beta}^\dagger \Psi_{\beta} + \langle \Psi_{\beta}^\dagger\Psi_{\beta} \rangle \Psi_{\alpha}^\dagger \Psi_{\alpha}.
    \end{aligned}
\end{equation}
This expression contains three kinds of mean fields:
i) Hartree terms like $\langle \Psi_{\alpha}^\dagger \Psi_{\alpha}\rangle$, which correspond to the shift in energy due to the presence of another particle in a nearby energy band; ii) Fock fields terms like $\langle \Psi_{\alpha}^\dagger \Psi_{\beta} \rangle$, which correspond to spin flips and are usually neglected in BCS theory; 3) Bogoliubov pairing terms $\langle\Psi_{\alpha}^\dagger\Psi_{\beta}^\dagger\rangle$, which correspond to the energy associated with forming a pair.     

In the Fourier space, the wavefunction $\Psi_{\alpha}(x)$ decomposes as
\begin{equation}
\Psi_{\alpha}(x)= \frac{1}{\sqrt{L}}\sum_k c_{\alpha k} e^{i k x},
\end{equation} 
where $c_{k \alpha}$ is the annihilation operator of an electron in the subband $\alpha$ with a wavevector $k$ and where $L$ is the quantization length. The interaction Hamiltonian [Eq.~(\ref{Hinteraction})] together with the assumption [Eq.~(\ref{ansatz})] can then be rewritten upon performing the integral as Eq.~(\ref{HinteractionSQ}). Hence, the interaction strength $U = VL$ appearing in Eq.~(\ref{HinteractionSQ}) contains both the base interaction strength $V$ and the quantization length $L$.  

\bibliographystyle{apsrev4-2}
\bibliography{bibliography}

\begin{thebibliography}{58}
\bibitem [{Com()}]{Comment1}%
  \BibitemOpen
  \href@noop {} {}\bibinfo {note} {Note that $\alpha_R$ is independent of the
  effective masses: their values are only important to evaluate the spin-orbit
  coupling between bands far from the $\Gamma$ point}\BibitemShut {NoStop}%
\end{thebibliography}

\end{document}